\documentclass{hotnets22}

\usepackage{times}  
\usepackage{hyperref}

\hypersetup{pdfstartview=FitH,pdfpagelayout=SinglePage}

\begin{document}

\title{Exposed Buffer Architecture}
\author{\begin{tabular}{c}
Micah Beck \\
mbeck@utk.edu \\
University of Tennessee \\
June 2022
\end{tabular}}
\date{June 2022}

\maketitle

\begin{abstract}The Internet stack is not a complete description of the resources and services needed to implement distributed applications, as it only accounts for communication services and the protocols that are defined to deliver them. 
This paper presents an account of the current distributed application architecture using a formal model of strictly layered systems, meaning that services in any layer can only depend on services in the layer immediately below it. 
By mapping a more complete Internet-based application stack that includes necessary storage and processing resources to this formal model, we are able to apply the Hourglass Theorem in order to compare alternative approaches in terms of their "deployment scalability." In particular,
we contrast the current distributed application stack with \textit{Exposed Buffer Architecture}, which has a converged spanning layer that allows for less-than-complete communication connectivity (exposing lower layer topology), but which also offers weak storage and processing services. 
This comparison shows that Exposed Buffer Architecture can have deployment scalability greater than the current distributed application stack while also providing minimally requisite storage and processing services.
\end{abstract}

\section{Introduction}
\vspace{8pt}
Certain shared elements of information infrastructure have been instrumental in the success of the modern application environment.
Notable among these have been the POSIX kernel interface and the Internet Protocol Suite.
Increasingly adopted as de facto standards in academia and many industry niches, they have enabled portability of code and interoperability of communication.
Results have included the creation of public shared infrastructure and active open source software communities.
Today the focus of innovation is in distributed applications which blur the lines between the traditional information technology silos: storage, networking and processing.
Network innovators have long sought to apply the design discipline of layered systems that has been the basis of the Internet's growth and acceptance to encompass distributed storage and processing resources.
It has sometimes seemed that venerable Internet design principles ruled out such convergence.\cite{anderson2005overcoming}
In this paper we present Exposed Buffer Architecture, a framework for applying the principles of layered systems to the design of converged infrastructure for distributed applications.\cite{9356087}

Layering of system components is an idea that has been used in many different ways. 
In the most general sense, layering assumes that there exists a directed "used by" relation between components.
System components are grouped into sets that share some important commonality.
If these sets have a natural partial order (from higher to lower) then they can be thought of as being arranged as layers stacked upon one another. 
Then if the "used by" order is consistent with the natural partial order of the layers, the common properties of the layers characterize all "used by" paths of components.

In telecommunications the notion of layering can be traced back to the information-theoretic notion of encapsulated protocols.
A higher level communication protocol A is encapsulated within a lower level protocol B when the latter is used as the channel to carry the data transmitted by the former (in the sense of the Shannon-Weaver model).\cite{ShannonWeaver49}
This notion of encapsulated communication does not take account of any storage or processing resources used by the endpoints or to implement the channel.

A more general use of layering is found in the description of computer systems.
Modularity of hardware and software components is a form of abstraction.
In computer systems a higher level function can use a lower level one by making a request of it.
If it is possible to characterize the effect of making a request of the lower level function then it is possible to reason about the implementation of the higher level function without knowing the implementation of the lower level one.
The impact of a service request on the state of the system can result in the allocation and use of communication, storage, processing or other resources.

The Internet stack was designed in the style of encapsulated communication protocols, with each layer being used as the channel for the next higher layer.
However there is necessarily more to the implementation of the Internet than data transmission.
Intermediate nodes at each layer run algorithms (e.g. for routing or integrity checking) and perform buffering (e.g. for reordering or retransmission) that use storage and processing resources.
Applications make very general and sometimes intensive use of storage and processing resources.

The Internet stack does not model the services necessary to implement general data storage or processing functions. This means that providing access to such resources for the implementation of the stack is the responsibility of the operators of intermediate nodes and endpoints.
This gives them great autonomy, but it generally excludes access to those services from analyses of the layered design of the Internet.
If not included in the model, the storage and processing resources needed to implement the layers of the stack are implemented in ad hoc manner.

The original design of the Internet excluded general access to storage and processing resources from the services it offers.
This was seen as necessary to create a network that could grow to a global scale while remaining usable (stable and predictable).
The idea that synchronous end-to-end unicast protocols and the client-server model of wide area services was sufficient to meet Internet application demands was widely accepted as a simplifying assumption.

This paper makes use of a formal model of layered systems to analyze the scaling properties of the current distributed application stack.
The resources modeled in this stack include not only communication but also storage and processing.
In order to evaluate how the use of these resources impacts the implementation of services provided by intermediate nodes and endpoints, the services that provide access to them must be included as explicit system components.
The resulting stack is capable of representing the Internet stack as a subset but also it can model layered decomposition of a more general distributed system architecture.

\section{A More General Model}
\vspace{8pt}

In our model there are only three layers, and components in the upper layer (consisting of applications of the middle layer services) cannot access the services of the lower layer (consisting of implementations of the middle layer services).
Storage and processing are implemented by devices, and so must be present in the lower layer of the model.
Thus, in order for applications to make use of storage and processing there must be a {\em "used by" path} from every component in the middle and upper layers to some storage and processing resources in the lower layer.

We map the current Internet stack to our three-layered model by assigning the Network layer of the former to middle layer of the latter.
All layers below the Network layer of the Internet stack are grouped into the lower layer of the three-layered model (which we will henceforth refer to as the "lower bell") and all layers above it to the upper layer (the "upper bell").
We refer to the middle layer of the three-layer model as its  "waist".
Note that our formal model does not make any distinction between components implemented on intermediate nodes and those implemented only at endpoints.
Thus we cannot reach any conclusions based on that distinction using this model.

We stress that the Internet stack does not explicitly include as components the storage and processing resources used to implement each layer.
In order to compare the current distributed application stack to alternatives that are designed to include such resources, we add to the middle layer services that make those resources available to applications.
For example, to describe the current state of distributed applications we include endpoint operating system services as well as those of network caches, replicated servers and other tools used by implementers.
This represents an expansion of the communication stack to include all (or at least more of) the resources used in the implementation of network services and applications.

This expanded description of the current distributed application stack is not a design.
It is rather a reflection of the reality of current implementation strategies.
As we will show, this model allows us to compare the current implementation to alternatives that have greater capacity for interoperability, generality and "deployment scalability".

The term "scalability" can be applied to many different dimensions of any system.
Networks and distributed systems can scale to consist of a large number of nodes.
A number of different performance metrics can scale with the application of increased resources.
In applied systems there is a notion that the design of a service or protocol can lend itself to widespread voluntary adoption by a community that crosses a variety of different boundaries such as geography, administrative domain and through the evolution of underlying implementation technologies.
We term this (admittedly somewhat vaguely defined) property "deployment scalability". 
The POSIX kernel interface and the Internet Protocol Suite are two notable examples of designs that have exhibited high deployment scalability.

\section{The Internet Application Stack}
\vspace{8pt}

The Internet Protocol Stack is a layered system in which the system components are message passing protocols.
In order for any service or application to be an interoperable element of the Internet environment it must use the Internet Protocol Suite (IPS) as its only mechanism to communicate with other elements. 
David Clark termed the IPS the “spanning layer” and we adopt this terminology to characterize the waist of our three-layer formal model.\cite{NAP6062}

We say that the IPS is only part of the waist of our model because that model must also include the services used to give the upper bell access to storage and processing resources.
Because there is no single standard in this area there are many different such services.
For example the implementation of TCP uses the internal memory management facilities of the host operating system to implement buffer management.
On the other hand applications implemented by processes in a general purpose operating system use the memory management and storage services provided at user level.
Processing performed in the upper layer similarly makes use of a number of different thread and process mechanisms.

Thus the storage and processing components of the current execution environment's waist is sprawling and changes constantly.
However there is a subset that is relatively fixed, such as the interface offered to user processes by POSIX- or Windows-compliant operating systems.
Similarly, the internal services of a number of popular POSIX and Windows implementations are also fairly stable.
The services used by other components, such as the app environment for streaming appliances, are more volatile but for the purposes of analysis the most mature and stable ones could be included.
This situation still makes for a messy set of services that have no overall unifying design.

Someone accustomed to working in the restricted context of the Internet communication stack might remark that this hardly seems like a spanning layer at all.
After all, what has just been described is not a single standard - there are many options to choose from.
But the reader would do well to bear in mind that the IPS is also not completely uniform in its implementation. 
There are optional features or variable characteristics which clients must discover and detect.
Notable examples include variation in Maximum Transfer Unit of an intermediate node and the availability of multicast routing and forwarding services.

The Internet Engineering Task Force standards process has kept such variations to a minimum, but uniformity is not absolute.
Certainly it is reasonable to think that unrestricted expansion of options within the waist would have a negative effect on interoperability.
In fact, today there is little interoperability between applications and utilities that access storage and processing resources using different mechanisms.
It is arguable whether such lack of interoperability is unavoidable, but it is easy to see that the impact is a matter of degree.
We will discuss the idea of minimizing such heterogeneity in a later section.

Similarly, the IPS is not completely static.
Standards have evolved, such as the move to IPv6 and the updating of multicast routing protocols.
Here again the IETF has been conservative but change does occur.
The lack of interoperability is a matter of degree.

For now, we focus on a reasonably stable subset of the current waist layer of the distributed appliaction stack.
While any such environment must include the IPS as part of the spanning layer, we leave the choice of the storage and processing services to the reader.
Our analysis applies to any reasonable choice.

\section{The Hourglass Theorem}
\vspace{8pt}

In software systems and infrastructure it has long been understood that interoperability can be enabled through the adoption by the developer community of a small set of services that are used by all higher layers. 
Higher level services that meet the interface of this common service layer can then use any implementation of it. This enables freedom for developers of lower layers that support the common layer, potentially leading to many alternatives. It also enables freedom for developers of higher layers leading to many applications. This is often referred to as the “hourglass model” because the common layer at its “waist” is constrained to conform to an agreed-upon standard, the “lower bell” of implementations and the “upper bell” of applications are free to grow.

The design of successful waists for layered systems has focused on a number of characteristics, including simplicity, minimality, elegance, generality and efficiency. 
In some circumstances such a design may be constrained by the need to make use of existing lower layers for implementation and the need to support existing applications. 
The greatest design freedom is possible in situations where new platforms and new applications are being developed. In such cases it is the experience, skill and intuition of individuals or small groups of system architects that have crafted our most successful designs, in particular POSIX and the IPS.

There is a formal theory of deployment scalability in layered systems that can shed light on the implications of modeling functionality such as storage and processing in the waist. \cite{10.1145/3274770}
The central result of this theory, the Hourglass Theorem, in some cases leads to conclusions that correspond to the classical End-to-End Arguments.\cite{Saltzer:1984:EAS:357401.357402}
In other cases the Hourglass Theorem predicts improved deployment scalability as consequences of network design choices (such as lack of complete symmetric connectivity) that are different from those made by the original designers of the Internet. 

The formal theory is only relevant in describing the structure of its narrow definition of deployment scalability as a function of the logical strength of the waist.
Formal deployment scalability is only one element among many that may inform any particular network design decision, and other considerations may be determinative in any specific scenario. 
However we ask the reader to consider the claim that this theory is often relevant to the success or failure of alternative designs.

Logical strength can be expressed in terms of how many guarantees a service makes to its clients. 
For example, best effort IP datagram delivery is quite weak because it does not guarantee correct or ordered receipt of any datagram that is sent. 
On the other hand a TCP connection guarantees in-order delivery of data with a low probability of data corruption, so it is stronger in these respects. 
Adding new functions to a layer requires additional guarantees to specify them, and thus results in a stronger specification.

The Hourglass Theorem concerns how many different lower layer services can be used to support the implementation the waist and how many higher layers applications can be supported by it. 
The issue is not how many {\em have been} implemented, but how many {\em can be}. 
The idea is that the more possible lower layers there are, the more environments and systems can support the spanning layer (i.e. the larger the lower bell of the hourglass can be). 
The more possible higher layers there are, the more applications can be supported by it  (i.e. the larger the upper bell of the hourglass can be). 
In this theory the formal definition of deployment scalability is having a large upper and lower bells simultaneously.

Informally, the Hourglass Theorem states that increasing the strength of a spanning layer increases the set of possible applications (the upper bell) but decreases the set of possible implementations (the lower bell). Note that “increasing” and “decreasing” actually refer to non-strict set inclusion. Sometimes a change in the strength of a spanning layer can leave either bell unchanged.

\section{Local Reachability}
\vspace{8pt}

Like the public telephone system, the original Internet architecture was designed to assign to each network interface a unique address chosen from a uniform address space.
There was also an assumption that endpoints providing services would be accessible from any network node.
These assumptions still apply to a subset of the globally routed Internet, but dynamic assignment of IP addresses, network address translation, firewalls and BGP anycast have weakened them considerably.

We propose weakening the waist of the application stack by allowing the set of connections directly supported by it to be determined locally.
Similarly, addresses can also be assigned locally.
There is no notion of globally unique addresses in the waist, and no assumption of global reachability.

This approach allows the notion of local adjacency to be reflected in the communication services of the waist.
The weaker the assumption of reachability, the larger the set of possible implementations.
The waist still abstracts all the technologies in the lower layer of the stack, but it may not implement direct communication spanning localities within the global communication topology.

\begin{figure*}
\begin{center}
\includegraphics[width=6in]{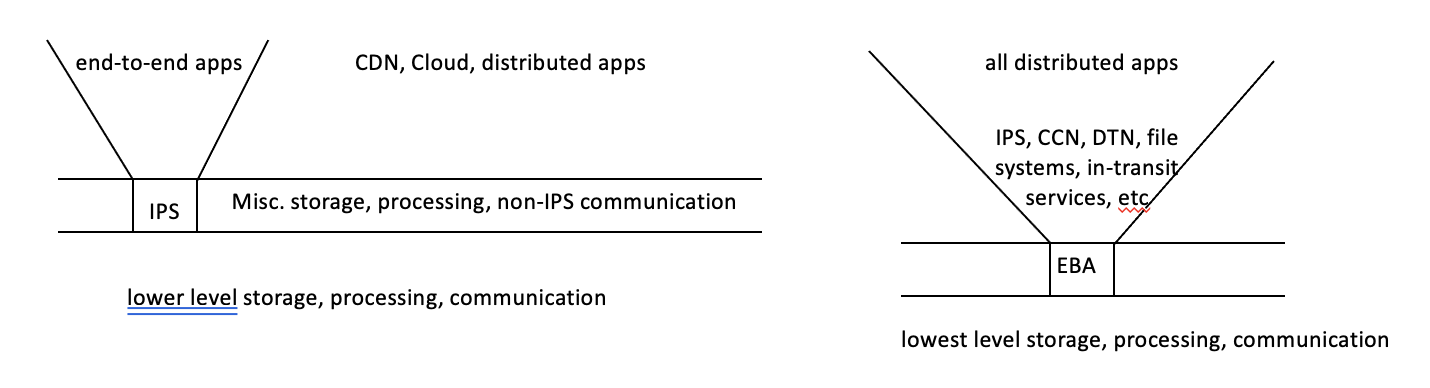}
\end{center}
\vspace{-24pt}
	\caption{The frame on the left illustrates the current distributed application stack with the Internet Protocol Stack is embedded at a waist that includes other ad hoc services.
	The frame on the right illustrates how the Exposed Buffer Architecture incorporates storage, networking and processing in a converged waist designed to be logically weak.\label{eba}} 
\end{figure*}

But wouldn't weakening the spanning layer in this way decrease the set of possible applications in the upper bell?
The answer to that question is no!
Global connectivity can be implemented in the upper bell.
In fact, the IPS can be implemented in the upper bell (see Figure \ref{eba}).
But because global routing would no longer be part of the spanning layer, alternative mechanisms could also be implemented.

Local reachability in the spanning layer is similar in effect to  placing the spanning layer at the Link layer of the Internet stack.
This general design was discussed by Clark in a 1997 paper when considering creating a stack in the shape of "a funnel placed on its end", with ATM at its narrow waist.\cite{NAP6062}
Early design discussions for the Global Environment for Network Innovation also considered moving the spanning layer to the Link layer.
Not requiring all applications to implement global communication in the same way, or at all, actually enables  a larger class of possible applications. 

An important difference between these proposals and ours is that those other designs did not consider including storage and processing resources as services of the spanning layer.
However instead of adapting an existing local area networking protocol as a standard (e.g. ATM or Ethernet), we might choose to design something more generic.
The Hourglass Theorem tells us that the choice that will have the greatest deployment scalability is the weakest one that is still capable of implementing necessary applications in the upper layer.

In this context logical weakness does not mean ruling out large connected subnetworks.
It simply means that the there is no minimum degree of reachability required.
it is acceptable if a node can only communicate with a small set of physically and/or trust adjacent nodes.
It is also acceptable for a node to have connectivity that is intermittent as long as it can always eventually connect to another node (as is the case in mobile ad hoc networks).

In this context, Content Centric \cite{DBLP:conf/acmicn/2019} and Delay Tolerant Networking \cite{10.1145/863955.863960}, as well as other mechanisms for more universal connectivity, could coexist with the IPS in the upper layer.
IPv4 and IPv6 currently coexist as mechanisms to achieve global reachability through translation mechanisms that bridge the two protocols.
Communication interoperability could be maintained through some combination of community standards and translation mechanisms.
Howver it is likely that communication interoperability between some highly divergent application categories is not required.
Heterogeneity and isolation are facts of life in the global distributed application environment that should be effectively modeled by infrastructure and then managed by policy.

The idea of a distributed application stack with a waist that only models local reachability may seem foreign to some in the networking community.
Distributed applications have been layered on top of Internet connectivity for so long that global reachability may seem definitional. 
In our proposal, the waist can be thought of as a standard node operating system or virtual machine that supports data transfer within a locality.
Interoperability then means the ability to manage data storage, processing and transfer within such localities.
From the point of view of all of these resources, the global distributed application environment is a composition of localities (homage to "a network of networks").

Eliminating the assumption of topology hiding allows services above the waist to not only take account of connectivity topology but also trust relationships and resource provisioning within network localities.
These can still be hidden from applications by utilities in the upper layer, but this is a decision in the design of each service.
Limited deployment scalability may be a necessary cost for some application communities which have strong requirements for topological information or control.

\section{Exposed Buffer Architecture}
\vspace{8pt}

Up to this point we have concentrated on localizing and exposing the communication topology of the waist of the distributed application stack to make it weaker.
We claim that including storage and processing in the spanning layer enables the upper bell to build utilities that create globally connected services.
However, we have acknowledged that the de facto nature of the storage and processing components of the waist have resulted in it being diverse (rather than standardized) and logically strong (relying on high level services).
This has led to a lack of deployment scalability.

At the application level storage and processing are typically accessed through services that are very strong.
This is partly due to history: processing was originally implemented by the allocation of an entire physical processor to a single task, and storage was implemented using punched cards or other write-once media.
Most users of computer systems reason about modern systems as if they were simulations of these original devices and media.
However the devices that implement modern information system generally offer much weaker services that then have to be composed and managed to provide strong abstractions.

We propose Exposed Buffer Architecture as a means of designing a spanning layer that incorporates storage and processing but is much weaker than the current distributed application stack.\cite{9356087} 

\begin{itemize}

\item Minimal strength in the allocation of data persistence resources (memory and storage) allows the restriction of allocations to a maximum size and duration determined by the storage resource owner and to best effort reliability. Such extreme weakness erases the distinction between a storage block and a memory frame, making both instances of a more general “buffer” abstraction but with different specific characteristics. This can be thought of as analogous to the blurring of “intranet” and “wide area” connectivity due to the uniform adoption of the IP protocol.
    
\item Minimal strength in processing requires operations to be invoked on explicitly named buffers containing all necessary information. Thus the equivalent of a task data structure must be an explicit argument rather than being held by the operating system and accessed only indirectly. Operations must have resource caps defined by the processing resource owner, including memory and processor utilization during operation execution, and reliability is best effort. Thus the abstraction that is currently implemented as an operating system process may have to be created through the composition of such limited operations and the explicit management of intermediate execution state. The choice of which operations are available  is under the control of the processing resource owner and is implemented through local administrative operations. Operations may range in nature from simple data movement to generic computations and even time slices of virtual machine execution.

\item Minimal strength in communication means restricting data movement to buffers on nodes that are within a set of logically adjacent nodes determined by the owner of the communication resource. This enables choices ranging from a completely isolated system to one that communicates only within a small cluster of physically colocated nodes to nodes that communicate within larger local area networks which may have strong properties. Limitations such as Maximum Transmission Unit, available bandwidth and maximum latency are under the control of the owner of the communication resources and reliability is best effort. 

\end{itemize}

The picture that emerges is of a minimal Node OS supporting local area communication.
In this model the Node OS does not necessarily support processes, files or sustained reliable communication but these may be constructed using local processing resources if the capabilities of the node allow it. 
The minimal instantiation of such a node is capable of only simple tasks such as packet forwarding, but with increasing resource bounds more complex and higher layer functionality can be supported. 
Wide area services must be built up out of these standardized local resources.
An overlay form of this architecture was implemented as the Logistical Networking Stack based on the Internet Backplane Protocol.\cite{Beck02anend-to-end,10.1007/978-3-030-12385-7_48}

\section{Point-to-Multipoint Apps}
\vspace{8pt}

While the early Internet application development community embraced the hiding low level topology, there was a problem when it came to implementing point-to-multipoint applications. In order to implement point-to-multipoint applications on a global scale it is necessary to introduce storage embedded within the low level network topology to avoid repeated transfer of information along the same link. This task can be accomplished effectively only with knowledge of the low level topology, and so the restriction prevented effective implementation using the Internet’s Network layer.

One early solution to this problem was the expansion of the Internet’s Network layer with Internet Protocol (IP) multicast. Being part of the Network layer, IP multicast could make use of lower layer topology and its implementation made direct use of IP routing tables. 
A full discussion of the reasons that IP multicast has not been more widely adopted is beyond the scope of this paper, but one issue is that it is a synchronous datagram delivery protocol, with the only delay between send and receipt being due to transmission and forwarding. 
The Internet architecture does not include access to storage service within the network, so receipt of information at different times by multiple receivers could not be implemented.

Another approach to implementing point-to-multipoint applications is to use storage implemented at network endpoints and accessed via point-to-point Internet communication. This client-server approach allows the Internet to be used without reference to lower layer topology and it supports asynchronous access, which is more general and can be more widely used. However as traffic levels increase this approach results in concentration of traffic that can quickly overwhelm the network and processing resources of a single server. Effective distribution of load requires the use of distributed storage (mirror sites or caches) and direction of traffic that takes account of low level topology. The inability for endpoints connected only using the Internet to take account of topology initially led to pushing the responsibility for the choice of server to end users (“Choose a mirror site near you”) or to static configuration by network administrators (configuration of Web proxy caches).

Eventually this gave rise to solutions implemented at IP endpoints that did make use of lower layer network topology, bypassing the Network layer to access lower layer services directly. Such solutions suffered from a lack of interoperability and were difficult for enterprises to implement. They required a detailed understanding of the structure of lower level topology, differentiating important features that change slowly from the “noise” of less stable elements. The difficulty of building and operating such solutions gave rise to Content Delivery Networks that invested heavily in infrastructure and business relationships that gave them control over their own collection of endpoints and the edge networks that they served. This became a sustainable business model when acting as a carrier for client services that are profitable and can sustain the cost.

We can apply the Hourglass Theorem to the example of exposing lower layer network topology in the Internet’s thin waist. A spanning layer that only exposes specific facts about the lower layer topology is logically weaker than one that passes through unfettered access to those layers. An example would be  telling the end user a best effort estimate of the Autonomous System hop count to a particular network address. Comparing the weakness of those two alternatives, the Hourglass Theorem assigns greater deployment scalability to the weaker spanning layer.

We stress that this model is descriptive. Using it does not mean we are advocating adding this functionality to the communication stack. 
The spanning layer of the distributed application stack already includes them, and this model simply reflects that reality. It does not change the definition of the communication stack, which is a subset of the distributed application stack. 
This reflects the fact that current distributed application solutions, including CDNs and distributed Cloud data centers,  make ad hoc use of services that do include these functions, resulting in low deployment scalability.

\section{Conclusions}
\vspace{8pt}

The question that the Hourglass Theorem can address is: Can a stack that implements the same (or larger) upper bell have greater deployment scalability? In other words, can the lower bell be increased? The Hourglass Theorem says that if we can design a spanning layer that includes the same upper bell but is logically weaker, it will have greater deployment scalability. The “minimal sufficiency” design technique is to maximize deployment scalability by minimizing the strength of the spanning layer while still implementing a designated set of “necessary applications”. The Hourglass Theorem also tells us that it may be possible to achieve even greater deployment scalability if the set of necessary applications is reduced. This however requires a (possibly difficult) decision not to support some applications. 

An example of such a trade-off can be found in the support of applications which require exceptionally low communication latency. Requiring the spanning layer to be capable of providing this to every endpoint strengthens it, ruling out support by some lower layers and so decreasing deployment scalability. In order to maintain deployment scalability it may be necessary to exclude such applications, or at least to find implementation strategies that do not always require extremely low latency.

Can a weaker spanning layer be designed that models storage and processing and still supports all necessary applications? The current way that applications make use of low level services are through complex resource-specific silos that were not designed to minimize logical strength. Some applications use the application layer of the network and operating system stacks, including file system interfaces. These services have specifications that are quite strong, as they were designed to be user-friendly abstractions of lower level resources. Even the Network layer is a stronger abstraction of communication than similar local area services which need not support complete or symmetric connectivity.

The world has benefited greatly from ubiquitous deployment of shared communication infrastructure.
But some classes of distributed applications have an inherent requirement for distributed storage and processing resources in order to implement effective and efficient solutions.
Defining standards for public infrastructure that excluded the management of such resources served to keep the network stateless and passive for a while.
But ultimately it led application developers to build private infrastructure that includes these resources, available only to those customers who can pay a substantial price.
We propose to extend ubiquity through deployment scalability to a general class of distributed applications.
To do so we must apply the principles of layered architecture to the management of storage and processing resources.

\section*{Acknowledgements}
The author acknowledges the contributions of Terry Moore and Elaine Wenderholm to this publication.

\bibliographystyle{abbrv} 
\begin{small}
\bibliography{ebabib,moreeba,hglass,MASS,SYSTOR}
\end{small}

\end{document}